\documentclass[12pt,preprint]{aastex}
\usepackage{psfig}
\def\HI {H\kern0.1em{\sc i}} 

\def\radm {rad m$^{-2}$}

\begin{document}
\title{~~\\ ~~\\ VLBA Imaging Polarimetry of Active Galactic Nuclei -- An Automated Approach}
\shorttitle{VIPS Pilot}
\shortauthors{Taylor et al.}
\author{G. B. Taylor\altaffilmark{1,2}, C. D. Fassnacht\altaffilmark{3}, 
L. O. Sjouwerman\altaffilmark{2}, S. T. Myers\altaffilmark{2}, 
J. S. Ulvestad\altaffilmark{2}, R. C. Walker\altaffilmark{2}, 
E. B. Fomalont\altaffilmark{2}, T. J. Pearson\altaffilmark{4}, 
A. C. S. Readhead\altaffilmark{4}, N. Gehrels\altaffilmark{5}, and
P. F. Michelson\altaffilmark{6}
}
\altaffiltext{1}{Kavli Institute of Particle Astrophysics and Cosmology,
Menlo Park, CA 94025}
\altaffiltext{2}{National Radio Astronomy Observatory, P.O. Box O, Socorro, NM 87801}
\altaffiltext{3}{Department of Physics, University of California, Davis, 
1 Shields Avenue, Davis, CA 95616}
\altaffiltext{4}{Owens Valley Radio Observatory, California Institute of Technology, Pasadena, CA 91125}
\altaffiltext{5}{NASA Goddard Space Flight Center, Greenbelt, MD 20771}
\altaffiltext{6}{Department of Physics, Stanford University, 382 Via Pueblo Mall, Stanford, CA 94305}

\slugcomment{Accepted by the Astrophysical Journal (Supplement Series)}

\begin{abstract}

We present full polarization Very Long Baseline Array (VLBA)
observations at 5 GHz and 15 GHz of 24 compact active galactic nuclei
(AGN).  These sources were observed as part of a pilot project to
demonstrate the feasibility of conducting a large VLBI survey to
further our understanding of the physical properties and temporal
evolution of AGN jets.  The sample is drawn from the Cosmic Lens
All-Sky Survey (CLASS) where it overlaps with the Sloan Digital Sky
Survey at declinations north of 15$^\circ$. There are 2100 CLASS
sources brighter than 50 mJy at 8.4 GHz, of which we have chosen 24
for this pilot study.  All 24 sources were detected and imaged at 5
GHz with a typical dynamic range of 500:1, and 21 of 24 sources were
detected and imaged at 15 GHz.  Linear polarization was detected in 8
sources at both 5 and 15 GHz, allowing for the creation of Faraday
rotation measure (RM) images.  The core RMs for the sample were found
to have an average absolute value of 390 $\pm$ 100 \radm.  We also
present the discovery of a new Compact Symmetric Object, J08553+5751.
All data were processed automatically using pipelines created or
adapted for the survey.

\end{abstract}


\section{Introduction}

Among astronomical observations, Very Long Baseline Interferometry
(VLBI) imaging is unique in providing details about parsec-scale
structures in cosmologically distant objects.  The Pearson-Readhead
(PR; \nocite{pr88}Pearson \& Readhead 1988) survey revealed a wide
range in morphologies and paved the way towards motion and variability
studies.  Subsequent VLBI surveys such as the Caltech--Jodrell Bank
Flat spectrum survey (CJF; Taylor et al. 1996\nocite{tay96}), have
imaged $\sim$300 sources at 5 GHz, of which more than half (177) were
observed in full polarization \citep{pol03}.  The VLBA Calibrator
Survey (VCS; \nocite{bea02} Beasley et al. 2002) included $\sim$2000
sources, but at the expense of limited sensitivity and ($u,v$)
coverage since the driving goals were an astrometric grid of
phase-referencing calibrators, and not high-quality imaging.

A new survey of $\sim$1000 sources with full polarization
at 5 and 15 GHz would provide an unparalleled combination of size,
depth, and polarization information.  The large sample of high-quality
data sets, especially when combined with surveys at other wavelengths,
would provide an excellent resource for studying the physics,
environments, and evolution of active glactic nuclei (AGN).  The key
areas of study for a large survey are:

{$\bullet$ \bf Gamma-ray AGNs:} This new Very Long Baseline Array
(VLBA) survey, combined with the survey data from the 
Gamma-Ray Large Area Space Telescope (GLAST) 
mission \citep{geh99}, should revolutionize our understanding of AGN
jet physics.  After launch in 2007, GLAST will perform an all-sky
survey 30 times more sensitive than that of EGRET.  The number of AGN
expected to be detected is $\sim$3000--4500, although variability, especially
in the gamma-ray properties, may strongly influence these estimates
\citep{ver04}.  Correlations with milliarcsecond polarimetry are key
to identifying and understanding the gamma-ray sources.  Extrapolating
from the correlation between 5 GHz radio luminosity and peak gamma-ray
emission \citep{mat97}, we expect that most of these objects will be compact,
flat-spectrum sources with 5 GHz flux densities over 30 mJy.  EGRET
results and theoretical studies indicate that gamma-ray emission
emanates from jet outflows (e.g., Dermer \& Schlickeiser
1994\nocite{der94}), where high-energy non-thermal electrons
accelerated in shocks are thought to Compton-upscatter soft
photons. Milliarcsecond radio imaging can reveal the detailed
morphological structure of the core and outflow regions.  With a large
number of sources, correlations between radio morphology and gamma-ray
spectral properties will be determined with high significance, thus
providing constraints on models for baryonic acceleration and the
microphysics of electron-proton coupling.  In addition, the improved
error box of GLAST ($\sim$arcminutes) combined with the VLBA imaging
may solve the mystery of the unidentified high-latitude EGRET sources.

{$\bullet$ \bf Jet Magnetic Fields:} Recently, Pollack et al.\ (2003)
\nocite{pol03} have found a tendency for the parsec-scale cores of
quasars to have a magnetic field aligned in the direction of the jet,
which can constrain jet-launching/collimation models (e.g., Meier,
Koide, \& Uchida 2001\nocite{mei01}).  Similar observations of BL Lacs
have suggested that their magnetic fields are oriented perpendicular
to the jets (e.g., Gabuzda et al.\ 2000\nocite{gab00}).
Unfortunately, the number of BL Lacs and galaxies with good
polarization data are too small for a meaningful comparison between
different classes of objects.  The 1000-source survey will address
this problem by providing $\geq$ 100 objects in each class.

{$\bullet$ \bf AGN Environments:}
The survey data will be used to study propagation effects local to the
jet, including free-free absorption in sources with counterjets and
Faraday rotation measures (RMs) towards all sources with detected
polarization.  Typical quasars have RMs which are surprisingly large
($>$1000 rad m$^{-2}$; Taylor 1998) and time-variable (Zavala \&
Taylor 2001). Knowing the RM is also critical in order to make sense
of the jet properties as otherwise the RM will tend to ``smear-out''
intrinsic relationships.

{$\bullet$ \bf AGN Evolution:}
Compact Symmetric Objects (CSOs), by virtue of their small sizes ($<$1
kpc), are thought to be examples of young ($10^3 - 10^4$ yr) radio
sources that may evolve to become eventually the well-known FRII-type
radio sources (Readhead et al.\ 1996).  Only small numbers of CSOs are
known ($\sim$40; Peck \& Taylor 2000), limiting our ability to study
radio sources at early times in their evolutionary history.  A large
VLBA survey would approximately double the number of confirmed
CSOs.  


Here we present the results from a feasibility study for a possible
VLBI Imaging and Polarimetry Survey (VIPS).  While not a complete
sample in itself, this pilot study of 24 AGN was chosen to be a
representative selection of strong and weak sources and presents a
wealth of data on this modest sample.  This includes one new Compact
Symmetric Object (J08553+5751) and the first Faraday rotation
measure observations of a sample of faint sources.  We have
also developed and made available 
new pipelines for calibration and imaging that allow rapid reduction
of the VLBA data to finished products.  Throughout this discussion, we
assume H$_{0}$=70 km s$^{-1}$ Mpc$^{-1}$, $\Omega_M$ = 0.27, and
$\Omega_{\Lambda}$= 0.73.

\section{Sample Definition}

To facilitate multi-wavelength science, all targets lie within the
$\pi$\ sr region of the North Galactic Cap and the equatorial strips
to be covered by the Sloan Digital Sky Survey (SDSS; \nocite{aba04,
aba05} Abazajian et al. 2004, 2005).  Thus multi-color imaging (UV to
near-IR) will be available for almost all targets, and spectroscopic
information will be present for the optically bright targets (perhaps
33\%). For optically faint sources, photometric redshifts may be
estimated from the $ugriz$ imaging data \citep{wei04}.  In this region
there are 2100 flat-spectrum, compact (size $<$200 mas), and
relatively bright ($S_{8.5 GHz} > 50$ mJy) sources above 15$^\circ$
declination in Cosmic Lens All-Sky Survey (CLASS) \citep{mye03}, a VLA
survey of $\sim$12,000 flat-spectrum radio sources at 200 mas
resolution.  From this number we have selected a sample of 1000
sources never before imaged at high dynanmic range with VLBI.  The
declination limit of 15$^\circ$ is necessary to insure uniformly good
radio imaging quality.  The lower limit of 50 mJy has been chosen in
order to have sufficient SNR ($>$1.5 estimated at 15 GHz) on all
compact sources with a spectral index $>-0.5$ to permit
self-calibration within the coherence time at both 5 and 15 GHz, where
the spectral index, $\alpha$, is defined as $S_\nu \propto
\nu^\alpha$.  Heavily resolved sources (sizes $>$ 20 mas) may still be
undetectable, especially at 15 GHz where 25 mJy is needed on scales
$<$20 mas for self-calibration.  We note that there is an implicit
avoidance of the galactic plane in our selection since CLASS was
selected to have $|b| > 10^\circ$.  CLASS also had a spectral
selection of $\alpha > -0.5$ from the parent Green Bank Survey (GB6 --
\nocite{gre96} Gregory et al.\ 1996) at 4.85 GHz and the NRAO VLA Sky
Survey (NVSS -- \nocite{con98} Condon et al. 1998) at 1.4 GHz.  A
similar spectral selection was employed for the CJF survey to obtain a
high success rate with VLBI imaging.

We note that there should be
$\sim$1000 GLAST sources found in the area covered by VIPS.
Since this is comparable to the number
of VIPS sample members in the same region, and compact flat-spectrum
radio emission is believed to be a strong predictor of gamma-ray emission
\citep{mat01, sow03}, a milliarcsecond
radio imaging survey of the VIPS sample may be important for the 
scientific interpretation of the GLAST all-sky survey (as discussed
in \S 1).  


Here we present observations of 24 sources (see Table 1) selected
within the first and second data release areas of SDSS, and to be
representative of the VLBA VIPS sample as a whole.  The integrated
flux densities of the 8.4 GHz VLA parent sample range from 52 mJy for
J08585+5552 to 850 mJy for J08546+5757 (JVAS 0850+581).  A total of 
8, 7, 5 and 4 sources were selected from the ranges 50--100, 100--200,
200--400, and $>$400 mJy.

\section{Observations and Analysis}

The observations were carried out at 5 and 15 GHz on 14 March 2004, 15
March 2004, 28 June 2004, and 18 August 2004, using the VLBA 
of the NRAO\footnote {The National Radio Astronomy Observatory is a facility of
the National Science Foundation operated under cooperative agreement
by Associated Universities, Inc.}.  Each observing session lasted for
12 hours.  The 5 and 15 GHz frequency bands were interleaved in time,
with 24 minutes scheduled on each target source at 5 GHz and 72
minutes on source at 15 GHz.  Allowing for telescope slews and other
overhead (6 minutes on average) the time needed for each source was
1.7 hours.  In order to improve ($u$,$v$) coverage and to allow for
rotation measure determinations, we spread the frequencies out over
the 500 MHz instantaneous tuning range of the VLBA as follows: 5 GHz
refers to 4607, 4677, 4992 and 5097 MHz at band center, while 15 GHz
refers to 14904, 14970, 15267, and 15366 MHz at band center.  Each
frequency was observed with 4 MHz bandwidth in each of right circular
and left circular polarization.  The only significant loss of data
were the loss of 98 minutes due to a false fire alarm at the Hancock,
NH VLBA station on 18 August.  Rain at some sites increased system
temperatures and reduced sensitivity, especially at 15 GHz, in all
four runs.

Amplitude calibration was derived using measurements of the system
temperatures and antenna gains at 4992 and 15366 MHz, which are close
to the continuum default frequencies of 4999 and 15369 MHz.
Fringe-fitting was performed with the AIPS task FRING on the strong
calibrator 3C~279.  Feed polarizations of the antennas were determined
at 5 and 15 GHz using the AIPS task LPCAL and the unpolarized source
OQ208.  An initial phase self-calibration was performed for each
source using a point source model -- no phase referencing to nearby
calibrators was used, as this would significantly reduce the observing
efficiency and is unnecessary for such strong sources.

Absolute electric vector position angle (EVPA) calibration was
determined by using the EVPA's of 3C~279, J0854+2006, and J1310+3220
as listed in the VLA Monitoring
Program\footnote{http://www.vla.nrao.edu/astro/calib/polar/}
\citep{tmy00}. Agreement in the EVPA correction determined from 
these calibrators was generally
better than 3 degrees for observations within a few days.
All observations of EVPAs for the target sources have been
corrected for Faraday rotation as determined from the 
observations.

For each source, the 15~GHz data were tapered to produce an image at
comparable resolution to the full resolution 5~GHz image.  For 
simplicity all sources were restored with a fixed beam size
of 3.2 $\times$ 1.6 mas in position angle 0$^\circ$ at each 
frequency. The two
images were then combined to generate a spectral index map.   It is
important to note that spectral index maps made from two datasets
with substantially different ($u,v$) coverages may suffer from
significant systematic errors, especially in
regions of extended emission.  Polarization images were likewise
made at matching resolutions in order to make 
rotation measure (RM) images.
All calibration and imaging was done semi-automatically using
pipelines written in AIPS \citep{gre03} and Difmap \citep{she97}.
These pipelines can be found at the VIPS 
web page\footnote{http://www.aoc.nrao.edu/$\sim$gtaylor/VIPS/}.

For 23 out of 24 sources the automatic imaging scripts were able to
produce an image with the expected noise and a good fit ($sigma$)
between the source model and the data.  The parameter $sigma$ is the
square root of the squared difference between the data and the model
divided by the individual variances implied by the visibility weights.
This is close to the square root of the reduced $\chi^2$, except that
a small change in the number of degrees of freedom due to amplitude
self-calibration has not been taken into account.  This simplification
is generally benign and will only result in a $\sim$10\% reduction of
$sigma$ for sources strong enough to amplitude self-calibrate (peak
flux density brighter than 0.3 Jy at 5 GHz or 0.5 Jy at 15 GHz).  The
value of $sigma$ indicates the agreement obtained between the model
and data in the self-calibration process with values near unity
indicating good agreement.  In one case a poor fit ($sigma > 1.4$) was
found to be due to bad data during 5 minutes from the Los Alamos, NM
station.  It is worth noting that in our inital automatic imaging
script we used a field size of 64 $\times$ 64 milliarcseconds, and
this resulted in poor fits for two sources with emission beyond the
edge of the field.  Judicial manual editing was done to solve the
first problem and imaging all sources with an initial field size of
128 $\times$ 128 milliarcsec then provided good fits for all 24
sources.  The final images have rms noise values within 30\% of the
predicted thermal noise.

The observing efficiency of 1.7 hours/source could be improved
to 1.5 hours/source 
without loss of sensitivity if the observing runs were extended
to 24 hours.  This could be done since the number of calibrator
scans is the same for a 24 hour run as it is for a 12 hour run.
The fraction of time spent on calibrators
for each of the 12 hour observing runs in the pilot project was 15\%.

\section{Results}

Figure 1 displays total intensity images for all sources in the pilot
project at 5 and 15 GHz.  Contours are drawn starting at 0.8 mJy beam$^{-1}$.
Further details can be found in Tables 2 and 3.  All 24 sources were
detected and imaged at 5 GHz.  At 15 GHz there was no
detection for J08553+5751 = JVAS 0851+580 (see \S 4.2.1),
J08585+5552, or J15406+5803.  Since no observations were carried
out with phase-referencing, sources with a flux density
less than 25 mJy at 15 GHz were too weak to self-calibrate.  The lack
of detections is due to a combination of moderately steep spectrum
sources and the lower sensitivity at 15 GHz within the coherence time
used for self-calibration of the phases.  The sensitivity limit for
detection was a flux density of 10 mJy at 5 GHz and 25
mJy at 15 GHz.

Linear polarization was detected in 8 of 24 sources at both 5 and 15 GHz,
and at 15 GHz only in the source J16542+3950.  In order to
increase the SNR in the polarization images the observations at 4607
and 4677 MHz were combined, as were those at 4992 and 5007 MHz, 14904 and 14970
MHz, and 15267 and 15366 MHz.   These 4 pairs were subsequently
combined to generate images of the rotation measure
by calculating the change in polarization angle with wavelength
squared on a pixel by pixel basis.  The results are shown in
Fig.~2, along with the magnetic field polarization vectors 
corrected for Faraday rotation.  Under the assumption that 
the radiation is optically thin synchrotron emission in a 
homogeneous field, the vectors shown indicate the projected
source magentic field orientation.

\subsection{\bf SDSS Magnitudes and Redshifts}

We have searched for optical counterparts for the sources in the SDSS
Data Release 3 \citep{aba05}, with a counterpart defined as a catalog
member falling within 1\arcsec\ of the radio position.  There are optical
matches for 22 of the 24 VIPS sources.  The detected sources have
$r$-band magnitudes between 17.5 and 22.1 (Table 1). The SDSS pipeline
produces a morphological classification of ``STAR'' for all but three
of the counterparts.  These sources are almost certainly quasars, and
we have designated them as ``Q'' in Table 1.  Furthermore, SDSS
redshifts are available for eight of the sources, all of which have
``STAR'' morphological classifications.  The spectra for these sources
are all typical quasar spectra, with broad emission lines of, e.g.,
CIV, CIII, MgII, or H$\beta$.  The redshifts range between $z = 0.5$
and $z = 2.0$.

\subsection{\bf Notes on Individual Sources}

\subsubsection{\bf J08553+5751} The morphology of this source 
revealed by our 5 GHz image (Fig.~1)
resembles that of a Compact Symmetric Object (CSO).  The identification of
0851+580 with a galaxy, and the lack of linear polarization, is also
consistent with a CSO classification, as is the lack of a detection
at 15 GHz. 

\subsubsection{\bf J14142+4554 }  This known CSO (JVAS 1412+461) 
\citep{pec00} was included in the sample as a check on the ability of
the survey to identify CSOs. In this regard the clear detection of two
steep spectrum lobes is quite satisfactory.  This source has been
associated with a galaxy of magnitude 19.9 and a redshift of 0.190 by
\cite{fal98}.  It has a bent northern lobe and an edge-brightened
southern lobe. \cite{gug05} find no detectable motions of the lobes
with an upper limit of 0.014 mas yr$^{-1}$, or 0.14~c.  This gives a
lower limit on the age of the radio source of 2030 yr.

\section{Discussion}

\subsection{Source Morphologies}

Of the 24 sources observed we have manually classified them
as 6 unresolved naked cores, 10 short jets (less than 10 mas), 4
long jets (more than 10 mas), 2
very bent jets, and 2 CSOs (see Table 2).  In Fig.~3 we plot the core
fraction, defined as the ratio of the peak to the integrated
intensity, versus the spectral index of the core component between 5
and 15 GHz.  As expected, the naked core and short jet sources 
have the flattest spectra (with one exception of a long jet source).  
Such sources are likely oriented 
at angles close to the line-of-sight and are prime candidates
for gamma-ray emission.

The two sources with highly bent jets have both steep spectra and a
low core fraction, possibly indicating that they are viewed at
moderately large angles to the line-of-sight.  In this case their
sharp bends are unlikely to be due to projection effects, and must be
intrinsically large.  For the one CSO detected at both 5 and 15 GHz,
the core fraction is overestimated since the peak in the image
corresponds to a hot spot, and not to the core.

\subsection{Evolution of AGN}

Compact Symmetric Objects (CSOs), by virtue of their small sizes ($<$1
kpc), are thought to be examples of young ($10^3 - 10^4$ yr) radio
sources that may evolve to eventually become the well-known FRII-type
radio sources \citep{rea96}.  Only small numbers of CSOs are known
($\sim$40; Peck \& Taylor 2000\nocite{pec00}), limiting our ability to
study radio sources at early times in their evolutionary history.

We present the discovery of a probable new CSO, J08553+5751.
Although this pilot sample is far from complete, and with only 23
sources the statistics are unreliable, the detection rate of 1/23 (not
counting the CSO J14142+4554 since it was deliberately selected) or
4\% is interesting.  \cite{tay03} found a detection rate of 2\% for
sources selected from the southern part of the VCS.  This is similar
to that obtained for the northern part of the VCS \citep{pec00}.  As
pointed out by \citet{pec00}, this detection rate is much lower than
that found in the PR sample of 7/65 (11\%), or 18/411 (4.4\%) found in
the combined PR and CJ samples.  We expect a slightly lower detection
rate because the VCS is comprised predominantly of flat spectrum
sources.  Nonetheless, the significantly lower detection rate found in
the VCS is probably the result of the reduced sensitivity and ($u,v$)
coverage compared to the PR and CJ surveys.  If the VIPS CSO detection
rate is well above 2\% then that indicates that the drop previously
seen is not the result of looking at the fainter end of the luminosity
function.  With a 4\% detection rate the VIPS survey would add 40 CSOs
to those known, roughly doubling the number of objects.  Assuming
Poisson statistics apply and we get 40 CSOs, then we should be 
able to determine the CSO fraction to an accuracy of $\sim$ 0.6\%.

\subsection{Magnetic Fields in the AGN Environment}

Typical quasars have rotation measures which are large
($>$1000 rad m$^{-2}$; Taylor 1998, Zavala \& Taylor 2003, 2004\nocite{tay98,zav03,zav04}) and time-variable \citep{zav01}. In the
VIPS pilot sample we find core RMs ranging from $-$22$\pm$45 \radm\ in
J16484+4104 to $-$931$\pm$20 \radm\ in J15457+5400 (Table 4).  The average
absolute value for the 8 sources shown in Fig.~2 is 390 $\pm$ 100 \radm.  This
value is somewhat less than the value of 640 \radm\ found by \cite{zav04}
for a sample of 40 strong AGN.   Presumably the Faraday screen
orginates in close proximity to the AGN jet, possibly created by
an interaction between the jet and the ambient gas.  Note that 
intrinsic RMs are larger by a factor of (1+z)$^2$.  The smaller 
RMs in the present sample could be the result of higher source redshifts, 
thinner Faraday screens, or small number statistics.  

Knowing the core RM value is also critical in order
to make sense of the jet properties as otherwise the RM will tend to
``smear-out'' intrinsic relationships.  With only 8 sources we 
do not presume to look for any trends in the magnetic field 
orientation.  

We note that the jets in 5 sources with extended polarization
detections all have low RMs (in the range $\pm$50 \radm), consistent
with results from \cite{zav04}.

\section{Conclusions}

We demonstrate that reliable, high dynamic range, total intensity VLBA
images can be obtained at 5 GHz for 100\% (24 of 24) sources with an
integrated flux density of over 50 mJy at 8.4 GHz, using a survey mode
with 1.7 hours total integration time per source.  For 21 of 24
sources (88\%) high dynamic range images were also obtained at 15 GHz,
allowing for the production of spectral index images.  We find a third
of the target sources can be imaged in linear polarization as well,
allowing us to determine Faraday rotation measures.  A pilot project
of 48 hours of observations yields a wealth of information for 24 AGN.
From this sample we report on the discovery of a new CSO, JVAS
J08553+5751.

Future observations of the complete VIPS sample of 1000 sources 
will provide an unparalleled combination of size,
depth, and polarization information.  The large sample of high-quality
data sets, especially when combined with surveys at other wavelengths,
will provide an excellent resource for studying the nature
of Gamma-ray loud AGN, the jet magnetic fields, the environs of
the AGN, and AGN evolution.

This
research has made use of the NASA/IPAC Extragalactic Database (NED)
which is operated by the Jet Propulsion Laboratory, Caltech, under
contract with NASA.

\clearpage

\begin{figure}
\figurenum{1}
\vspace{20.0cm}
\includegraphics{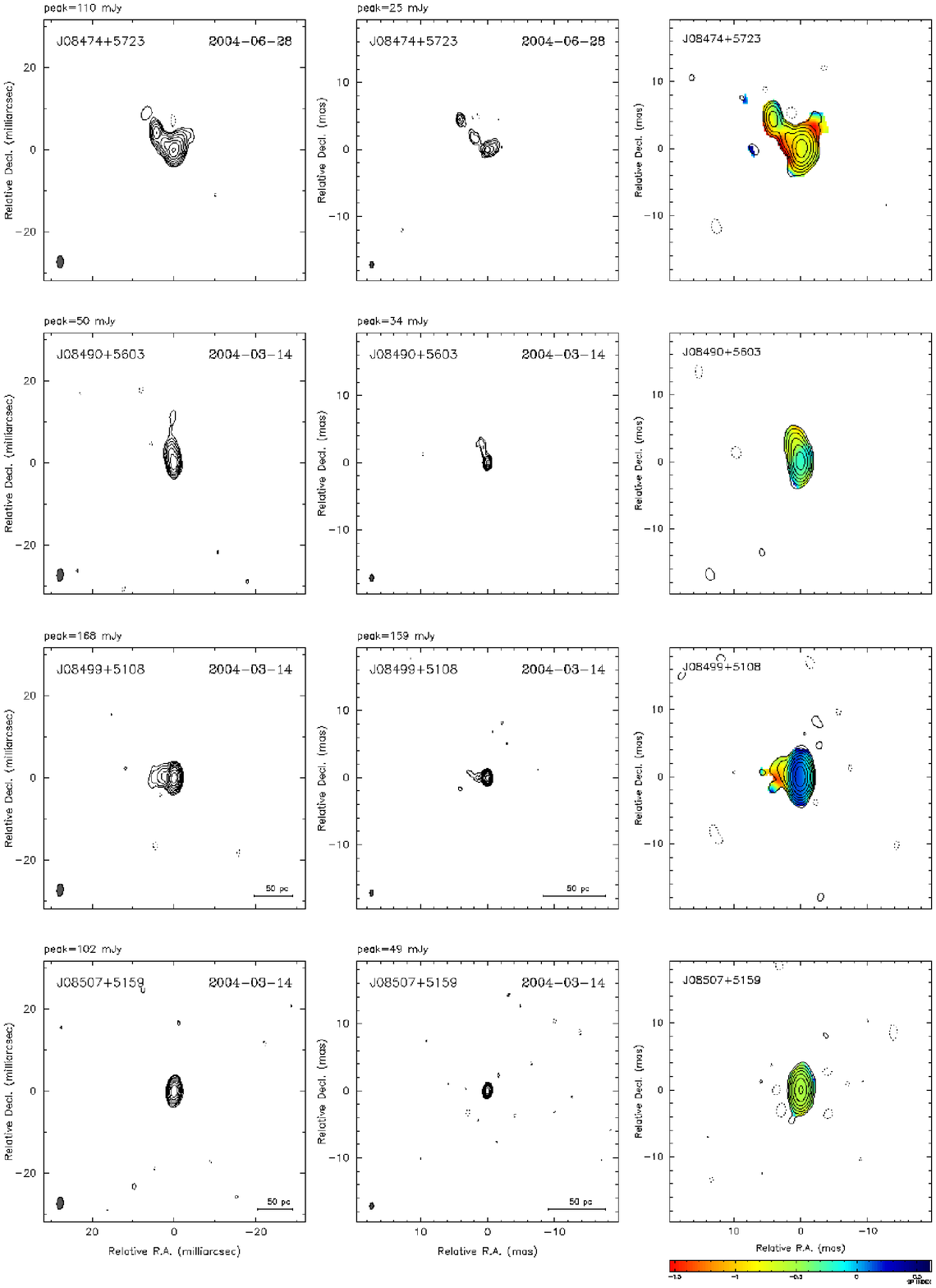}
\caption{ Total intensity contours of the VIPS pilot sources at 5 GHz (left),
  and 15 GHz (middle), along with spectral index (right).  The 15 GHz
  image and spectral index image cover the inner quarter of the 5 GHz field.
  Contours start at 0.8 mJy beam$^{-1}$ and increase by factors of 2.  
  Where redshifts are available, a
  50~pc scale is indicated.  The synthesized beam is shown in the bottom-left corner.  Image parameters are given in Table~2.  The spectral index image
is overlaid with 15 GHz contours at a fixed resolution.}
\end{figure}
\clearpage

\begin{figure}
\figurenum{1}
\vspace{20.4cm}
\includegraphics{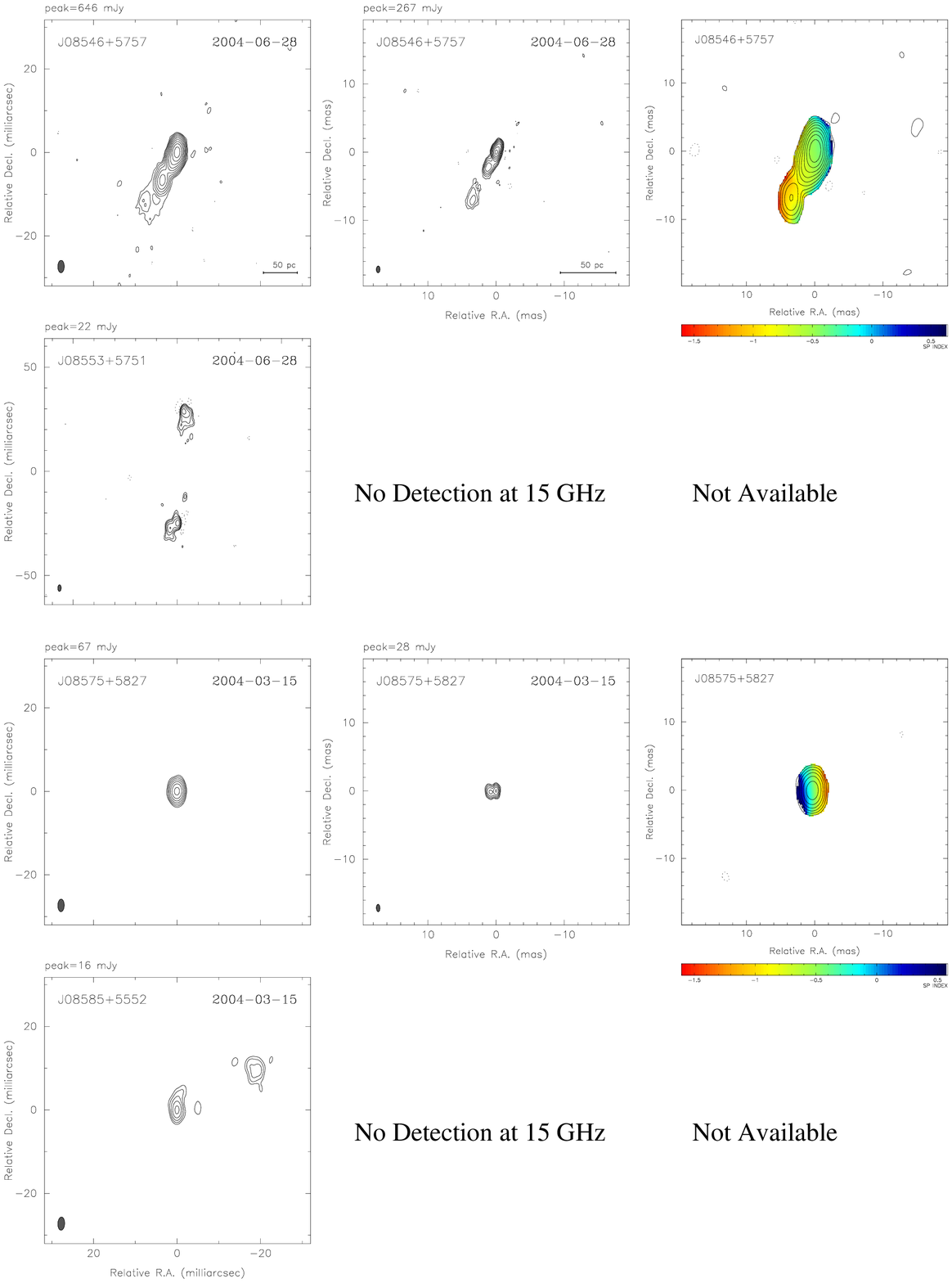}
\caption{ Continued.}
\end{figure}
\clearpage

\begin{figure}
\figurenum{1}
\vspace{20.4cm}
\includegraphics{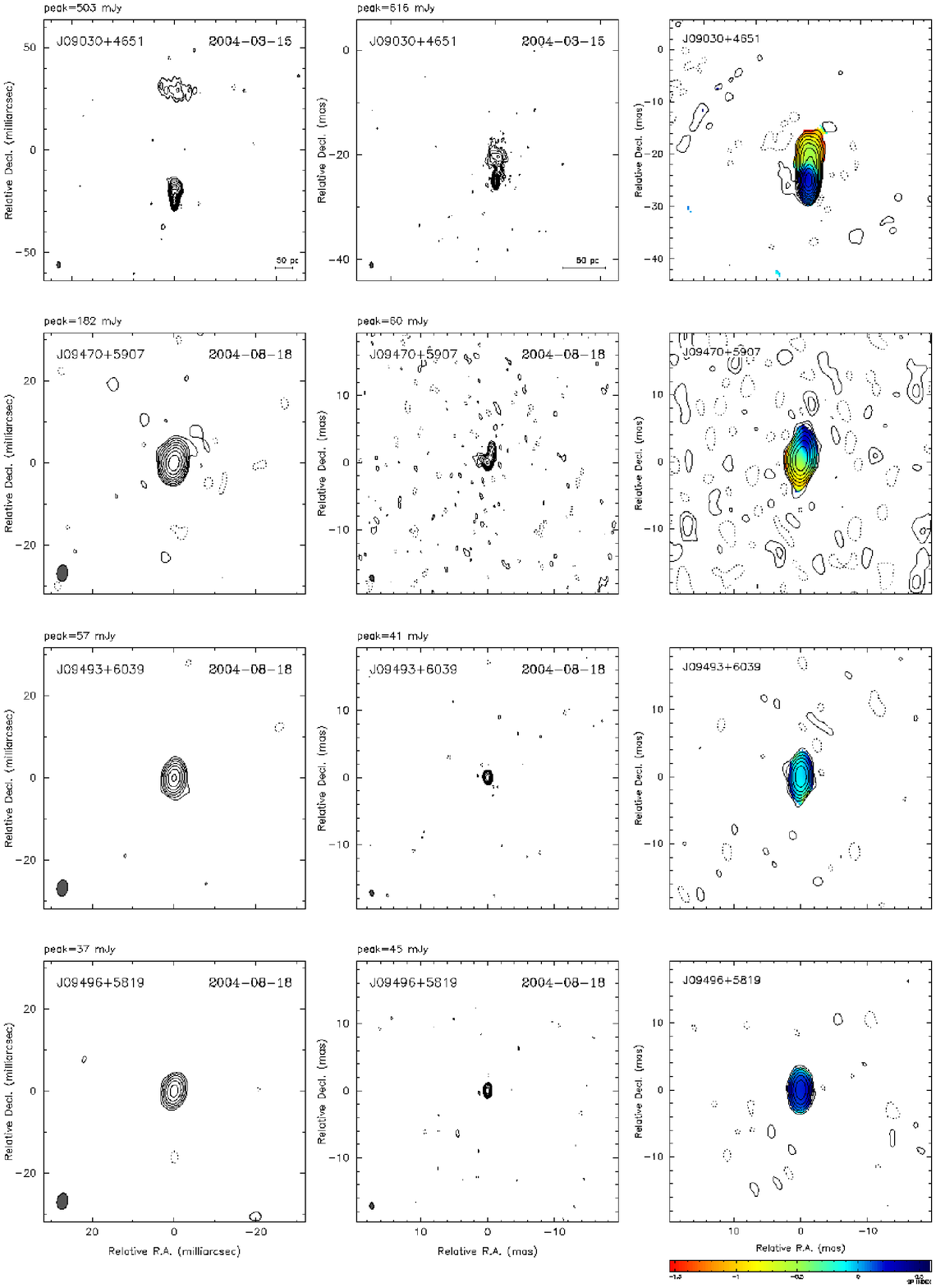}
\caption{ Continued.}
\end{figure}
\clearpage

\begin{figure}
\figurenum{1}
\vspace{20.4cm}
\includegraphics{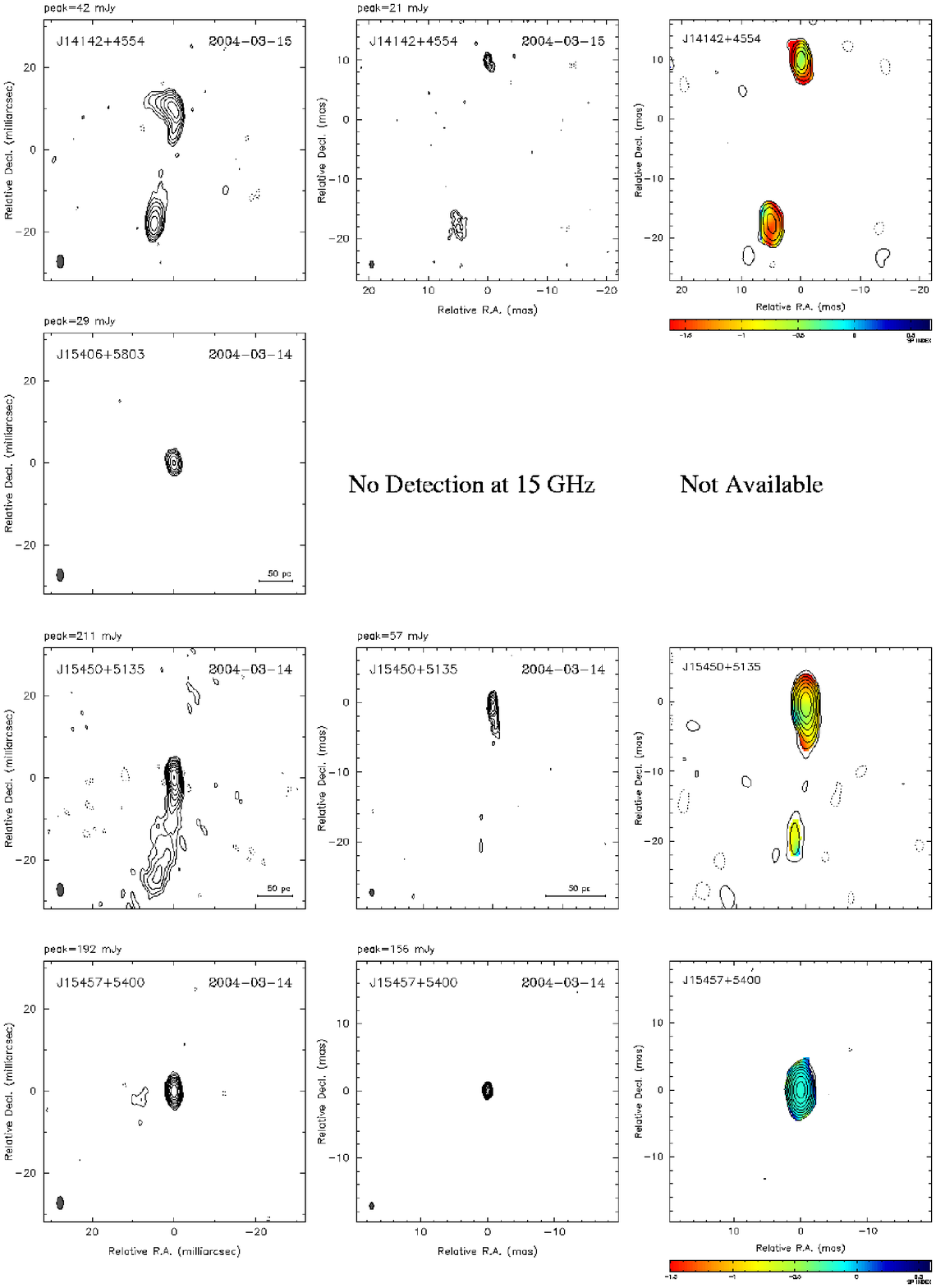}
\caption{ Continued.}
\end{figure}
\clearpage

\begin{figure}
\figurenum{1}
\vspace{20.4cm}
\includegraphics{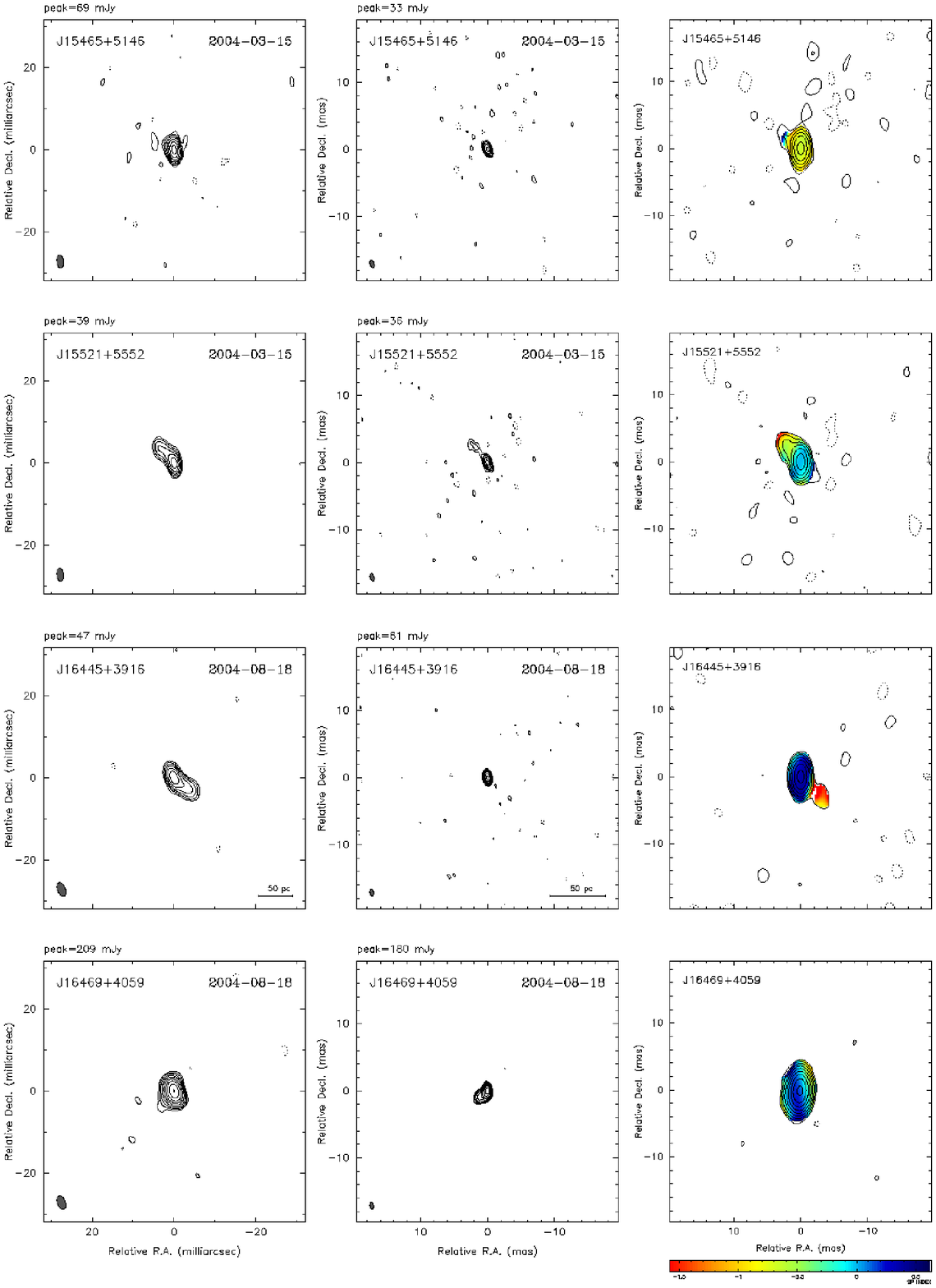}
\caption{ Continued.}
\end{figure}
\clearpage

\begin{figure}
\figurenum{1}
\vspace{20.4cm}
\includegraphics{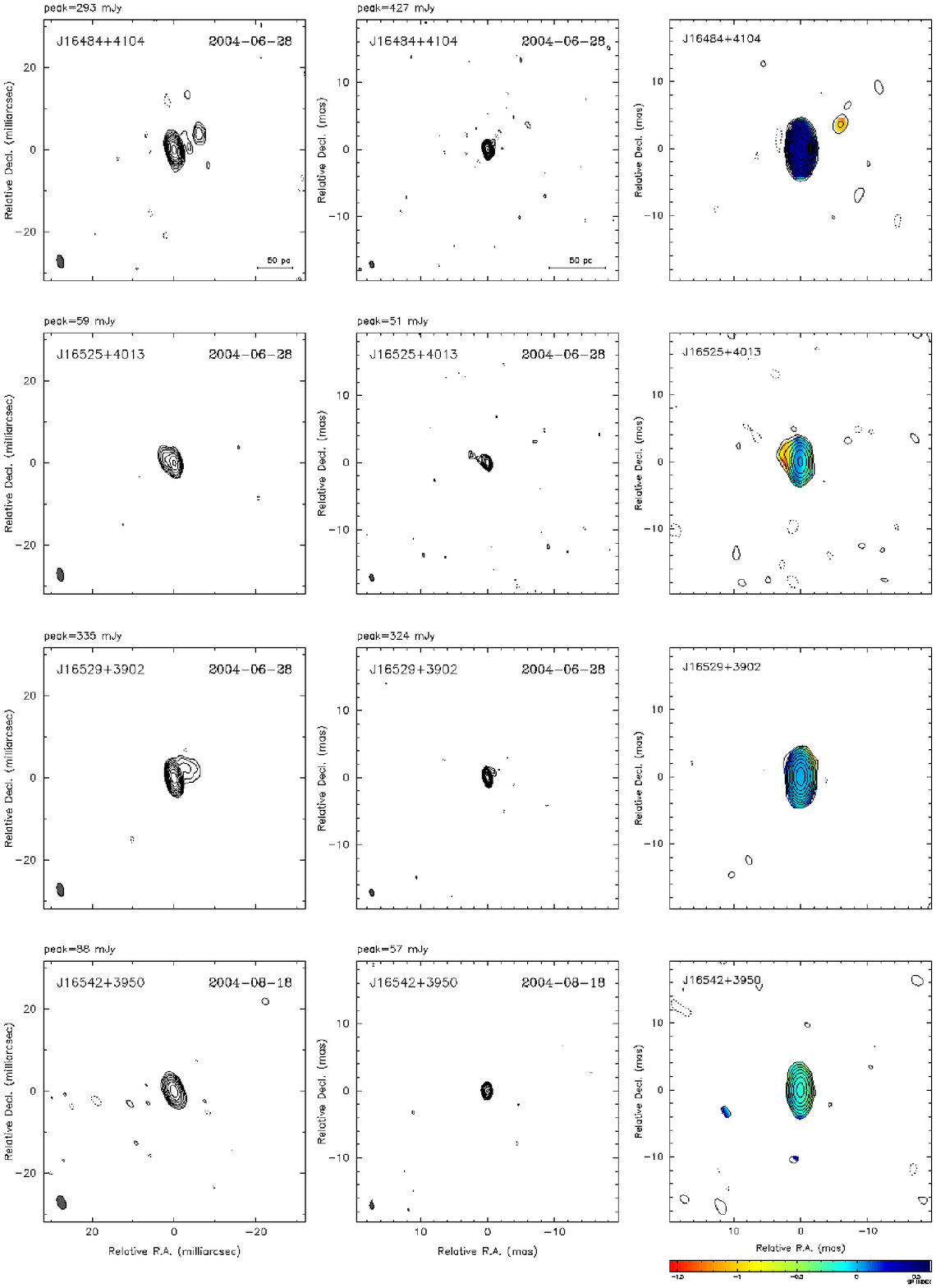}
\caption{ Continued.}
\end{figure}
\clearpage

\begin{figure}
\figurenum{2}
\vspace{20.4cm}
\includegraphics{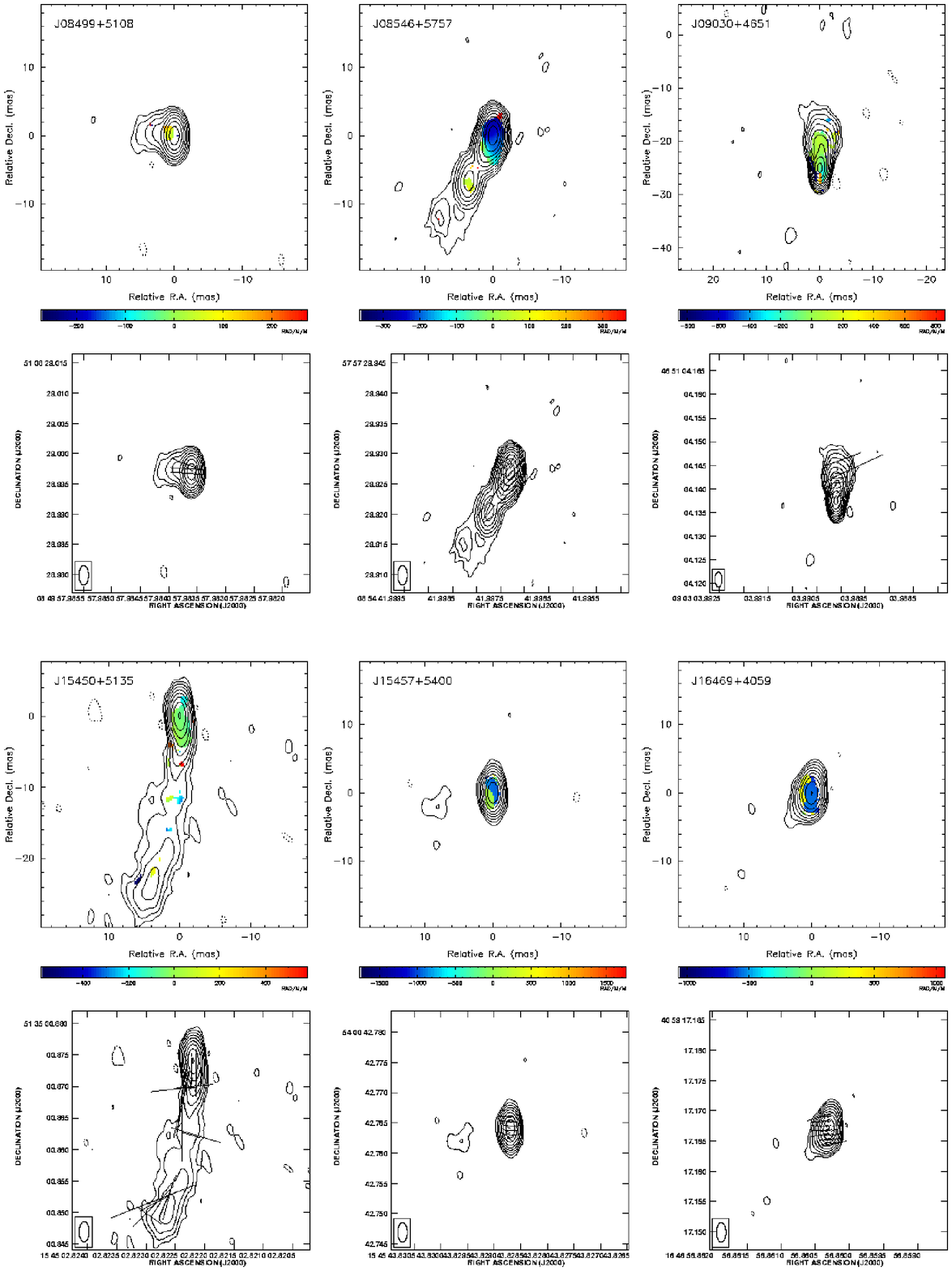}
\caption{ Polarization magnetic field (B) vectors at 5 GHz, along with
images of the rotation measure computed from polarization
angles measured at four frequencies.  The polarization angles have been
corrected by the observed RM, and the vector lengths are 
proportional to the fractional linear polarization.
Contours are at 5 GHz and drawn at the same levels as in Figure 1.}
\end{figure}
\clearpage

\begin{figure}
\figurenum{2}
\vspace{20.4cm}
\includegraphics{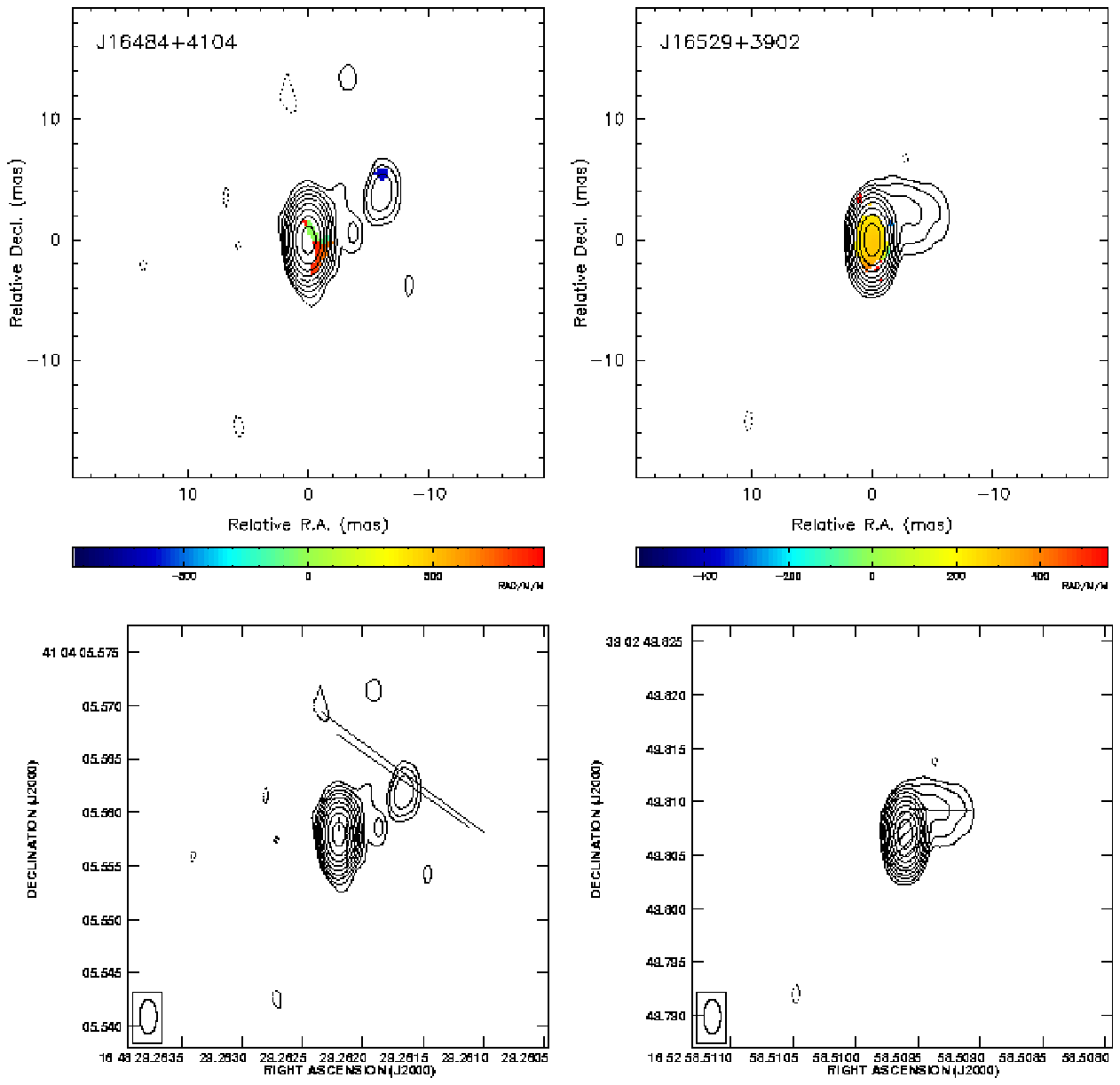}
\caption{ Continued.}
\end{figure}
\clearpage

\begin{figure}
\figurenum{3}
\vspace{20cm}
\includegraphics{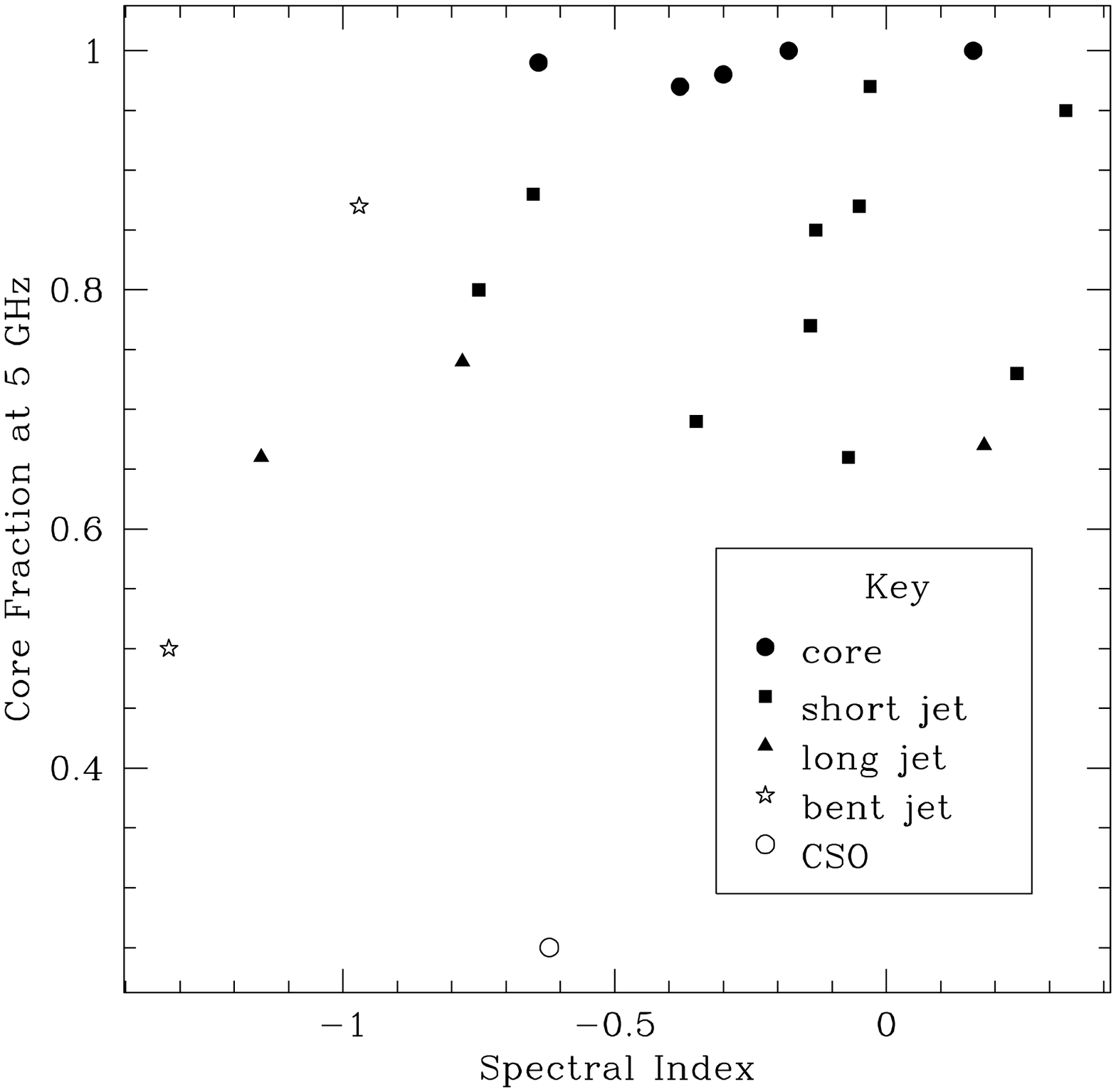}
\caption{ A scatter plot of the core fraction (peak intensity
divided by total intensity) versus spectral index computed between
5 and 15 GHz for 21 sources.  Sources have been classified as
either naked cores, short jets (length $<$ 10 mas), long 
jets (length $>$ 10 mas), highly bent jets, or Compact Symmetric
Objects (see Table 2).}
\end{figure}
\clearpage

\begin{deluxetable}{llllcllll}
\tabletypesize{\footnotesize}
\tablewidth{0pt}
\tablecolumns{9}
\tablecaption{VIPS Pilot Sample\label{tab1}}
\tablehead{\colhead{Source} & \colhead{Alternate} &\colhead{} & \colhead{} &
  \colhead{8 GHz} & \colhead{} & \colhead{} & \colhead{} & \colhead{} \\
  \colhead{Name} & \colhead{Name} & \colhead{RA} &  \colhead{Dec} & 
  \colhead{$S$} & \colhead{ID} & \colhead{$M_r$} & \colhead{$z$} & \colhead{$date$} \\
  \colhead{(1)} & \colhead{(2)} & \colhead{(3)} & \colhead{(4)} & 
   \colhead{(5)} &\colhead{(6)} &\colhead{(7)} &  \colhead{(8)} & \colhead{(9)} }
\startdata
J08474+5723 & 0843+575 & 08 47 28.0579 &57 23 38.349 & 240 & ... & ... & ... & C \\
J08490+5603 & &  08 49 00.8546 & 56 03 50.122 & 70 & Q & 21.267 & ... & A \\
J08499+5108 & &  08 49 57.9836 & 51 08 28.997  & 325 & Q & 18.194 & 0.584 & A \\
J08507+5159 & &  08 50 42.2482 & 51 59 11.674  & 112 &Q & 18.961 & 1.892 & A \\
J08546+5757 & 0850+581 & 08 54 41.9973 & 57 57 29.927 & 890 & Q & 17.739 & 1.318 & C \\
J08553+5751 & 0851+580&  08 55 21.3558 & 57 51 44.091 & 160 & G & 19.843 & ... & C \\
J08575+5827 & &  08 57 31.4511 & 58 27 22.538  & 88 & Q & 22.082 & ... & B \\
J08585+5552 & &  08 58 30.4786 & 55 52 41.302  & 52 & G & 18.473 & ... & B \\
J09030+4651 & 0859+470 & 09 03 03.9901030 & 46 51 04.13753 & 963 & Q & 18.989 & 1.47 & B \\ 
J09470+5907 & &  09 47 04.8631 & 59 07 41.467 & 70 & ... & ... & ... & D \\
J09493+6039 & &  09 49 20.2283 & 60 39 22.871 & 325 & Q & 19.917 & ... & D \\
J09496+5819 & &  09 49 39.8149 & 58 19 12.933 & 112 & Q &  21.308 & ... & D \\
J14142+4554 & 1412+461 & 14 14 14.8535061 & 45 54 48.65427 & 80 & G & 20.158 & ... & B \\
J15406+5803 & &  15 40 37.5779 & 58 03 34.398  & 71 & Q & 18.549 & 1.25 & A \\
J15450+5135 & 1543+517 &  15 45 02.8222 & 51 35 00.874  & 635 & Q & 17.552 & 1.93 & A \\
J15457+5400 & &  15 45 43.8287 & 54 00 42.764  & 145 & Q & 19.827 &  ... & A \\
J15465+5146 & &  15 46 33.6208 & 51 46 45.455 & 90 & Q & 19.943 & ... & B \\
J15521+5552 & &  15 52 10.8942 & 55 52 43.211 & 131 & Q & 20.688 & ... & B \\
J16445+3916 & 1642+393 & 16 44 34.4775 & 39 16 04.915 & 635 & Q & 19.808 & 1.583 & D \\
J16469+4059 & 1645+410 & 16 46 56.8603 & 40 59 17.167 & 71 & Q & 19.252 & ... & D \\
J16484+4104 & 1646+411 & 16 48 29.2622 & 41 04 05.558 & 210 & Q & 18.741 & 0.852 & C \\
J16525+4013 & &  16 52 33.2136 & 40 13 58.339 & 100 & Q  & 20.465 & ... & C \\
J16529+3902 & 1651+391 & 16 52 58.5096 & 39 02 49.807 & 330 & Q & 21.341 & ... & C \\
J16542+3950 & &  16 54 12.7223 & 39 50 05.681 & 145 & Q & 19.948 & ... & D \\
\enddata
\tablenotetext{~}{Notes - (1) J2000 source name in the IAU format HHMMd+DDMM; (2) Alternate name; (3) Right
ascension and (4) Declination in J2000 coordinates from CLASS \citep{mye03}, 
except for 1412+461 and 1543+517 taken from the International
Coordinate Reference Frame (ICRF; \nocite{ma98} Ma et al. 1998); (5) Flux density at 8.4 GHz
from CLASS; (6)
Optical host galaxy identification; (7) Optical magnitude in r band; (8)
spectroscopic redshift from SDSS; (9) Date of observation, A = 14 March 2004, 
B = 15 March 2004, C = 28 June 2004, and D = 18 August 2004.
}
\end{deluxetable}
\clearpage

\begin{deluxetable}{llrrrrrrr}
\tabletypesize{\footnotesize}
\tablewidth{0pt}
\tablecolumns{9}
\tablecaption{5 GHz VIPS Image Parameters\label{tab2}}
\tablehead{
  \colhead{} & \colhead{} & \colhead{} & \colhead{} & 
  \colhead{Peak Flux} & 
  \colhead{rms} & 
  \colhead{}  \\
  \colhead{Source} & 
  \colhead{Beam} & 
  \colhead{$\theta$} & 
  \colhead{Total Flux} &
  \colhead{(mJy} & 
  \colhead{(mJy} & 
  \colhead{Fit} & \colhead{Morph-} & \colhead{Core} \\
  \colhead{Name} & 
  \colhead{(mas)} & 
  \colhead{($^{\circ}$)} & 
  \colhead{(mJy)} &
  \colhead{beam$^{-1}$)} & 
  \colhead{beam$^{-1}$)} & 
  \colhead{$sigma$} & \colhead{ology$^*$} & \colhead{Fraction}
  }
\startdata
J08474+5723 & 2.99$\times$1.53 & $-$0.8 & 219.5& 110.0 & 0.21 & 0.998 & bj & 0.50 \\
J08490+5603 & 3.07$\times$1.50 & $-$4.0 & 72.6& 50.2 & 0.23 & 1.014 & sj & 0.69 \\
J08499+5108 & 3.03$\times$1.51 & $-$3.5 & 191.8& 167.5 & 0.19 & 1.009 & sj & 0.87 \\
J08507+5159 & 2.97$\times$1.50 & $-$3.8 & 103.5 & 102.3 & 0.24 & 1.064 & nc & 0.99 \\
J08546+5757 & 2.98$\times$1.54 & $-$0.6 & 875.3 & 645.8 & 0.23 & 0.979 & lj & 0.74 \\
J08553+5751 & 3.11$\times$1.61 & $-$0.7 & 44.4 & 26.2 & 0.21 & 1.015 & CSO & 0.59 \\
J08575+5827 & 2.99$\times$1.53 & $-$1.1 & 83.3 & 66.7 & 0.19 & 1.010 & sj & 0.80 \\
J08585+5552 & 3.14$\times$1.63 & $-$2.7 & 30.1 & 17.1 & 0.17 & 0.995 & lj & 0.57 \\
J09030+4651 & 3.11$\times$1.54 & 1.1 & 746.6 & 503.4 & 0.20 & 0.952 & lj & 0.67 \\
J09470+5907 & 3.96$\times$2.67 & $-$6.1 & 208.2 & 181.7 & 0.29 & 0.992 & bj & 0.87 \\
J09493+6039 & 3.97$\times$2.65 & $-$8.8 & 58.5 & 57.4 & 0.23 & 0.991 & nc & 0.98  \\
J09496+5819 & 3.95$\times$2.65 & $-$10.5 & 37.3 & 37.3 & 0.21 & 0.968 & nc & 1.00 \\
J14142+4554 & 3.36$\times$1.64 & $-$8.9 & 171.3 & 43.6 & 0.23 & 1.061 & CSO & 0.25 \\
J15406+5803 & 3.01$\times$1.54 & 1.4 & 35.2 & 28.7 & 0.18 & 0.990 & nc & 0.82 \\
J15450+5135 & 3.11$\times$1.56 & 3.1 & 318.7 & 211.4 & 0.27 & 1.102 & lj & 0.66  \\
J15457+5400 & 3.09$\times$1.52 & 0.5 & 192.5 & 192.4 & 0.21 & 1.036 & nc & 1.00 \\
J15465+5146 & 3.23$\times$1.59 & 4.9 & 78.2 & 69.2 & 0.23 & 1.026 & sj & 0.88 \\
J15521+5552 & 3.17$\times$1.56 & 5.6 & 59.9 & 39.4 & 0.20 & 1.000 & sj & 0.66 \\
J16445+3916 & 3.47$\times$1.92 & 21.6 & 64.2 & 46.7 &  0.23 & 0.983 & sj & 0.73 \\
J16469+4059 & 3.44$\times$1.93 & 20.4 & 247.5 & 209.3 & 0.21 & 0.985 & sj & 0.85  \\
J16484+4104 & 3.20$\times$1.55 & 11.5 & 307.7 & 291.5 & 0.25 & 1.086 & sj & 0.95 \\
J16525+4013 & 3.20$\times$1.53 & 9.9 & 76.6 & 59.2 & 0.21 & 0.992 & sj & 0.77 \\
J16529+3902 & 3.26$\times$1.54 & 10.7 & 345.0 & 334.5 & 0.19 & 0.912 & sj & 0.97 \\
J16542+3950 & 3.48$\times$1.93 & 20.6 & 91.4 & 88.5 & 0.24 & 1.017 & nc & 0.97 \\
\enddata
\tablenotetext{*}{
  Source morphology is either 'nc': naked core; 'sj': short-jet;
'lj': long-jet; 'bj': bent jet; or CSO: Compact Symmetric Object.
}
\end{deluxetable}

\begin{deluxetable}{llrrrrrrr}
\tabletypesize{\footnotesize}
\tablewidth{0pt}
\tablecolumns{9}
\tablecaption{15 GHz VIPS Image Parameters\label{tab3}}
\tablehead{
  \colhead{} & \colhead{} & \colhead{} & \colhead{} & 
  \colhead{Peak Flux} & 
  \colhead{rms} & 
  \colhead{} & \colhead{Spectral} \\
  \colhead{Source} & 
  \colhead{Beam} & 
  \colhead{$\theta$} & 
  \colhead{Total Flux} &
  \colhead{(mJy} & 
  \colhead{(mJy} & 
  \colhead{Fit} & \colhead{Index} & \colhead{Core} \\
  \colhead{Name} & 
  \colhead{(mas)} & 
  \colhead{($^{\circ}$)} & 
  \colhead{(mJy)} &
  \colhead{beam$^{-1}$)} & 
  \colhead{beam$^{-1}$)} & 
  \colhead{$sigma$} & \colhead{$\alpha$} & \colhead{Fraction}
  }
\startdata
J08474+5723 & 0.98$\times$0.52 & $-$2.2 & 66.9 & 24.6 &  0.19 & 0.972 & $-$1.32 & 0.37 \\ 
J08490+5603 & 1.00$\times$0.53 & $-$1.9 & 44.6 & 33.6 &  0.18 & 0.979 & $-$0.35 & 0.75 \\ 
J08499+5108 & 0.98$\times$0.51 & $-$2.8 & 198.2 & 159.0 &  0.19 & 0.981 & $-$0.05 & 0.80 \\ 
J08507+5159 & 0.98$\times$0.52 & $-$4.3 & 54.3 & 49.2 &  0.21 & 0.996 & $-$0.64 &  0.91 \\ 
J08546+5757 & 0.97$\times$0.53 & $-$1.0 & 495.5 & 266.9 &  0.21 & 1.014 & $-$0.78 & 0.54 \\ 
J08553+5751 & \\
J08575+5827 & 1.06$\times$0.52 & 1.7 & 52.5 & 28.4 &  0.16 & 0.972 &  $-$0.75 & 0.54 \\ 
J08585+5552 & \\
J09030+4651 & 1.10$\times$0.52 & 2.4 & 748.6 & 616.1 &  0.18 & 0.922 & 0.18 & 0.82 \\ 
J09470+5907 & 0.98$\times$0.53 & 7.4 & 133.2 & 59.9 &  0.31 & 1.019 & $-$0.97 &  0.45 \\ 
J09493+6039 & 0.94$\times$0.52 & 9.7 & 53.9 & 40.9 &  0.22 & 0.975 & $-$0.30 &  0.76 \\ 
J09496+5819 & 1.00$\times$0.53 & 1.8 & 45.0 & 45.0 &  0.23 & 0.975 & 0.16 & 1.00 \\ 
J14142+4554 & 1.21$\times$0.60 & 0.9 & 34.6 & 21.6 &  0.24 & 0.950 & $-$0.62 &  0.62 \\ 
J15406+5803 & \\
J15450+5135 & 1.04$\times$0.54 & 1.4 & 116.2 & 56.9 &  0.23 & 1.024 & $-$1.15 & 0.49 \\ 
J15457+5400 & 1.02$\times$0.53 & 0.3 & 156.6 & 155.9 &  0.18 & 0.993 & $-$0.18 & 1.00 \\ 
J15465+5146 & 1.18$\times$0.59 & 16.0 & 33.1 & 33.0 &  0.25 & 0.995 & $-$0.65 & 1.00 \\ 
J15521+5552 & 1.16$\times$0.59 & 17.0 & 42.6 & 36.4 &  0.27 & 1.003 & $-$0.07 & 0.80 \\ 
J16445+3916 & 1.05$\times$0.52 & 8.5 & 65.0 & 61.4 &  0.25 & 0.999 & 0.24 & 0.94 \\ 
J16469+4059 & 1.05$\times$0.52 & 8.4 & 265.1 & 179.5 &  0.17 & 0.983 & $-$0.13 & 0.68 \\ 
J16484+4104 & 1.04$\times$0.51 & 12.3 & 452.1 & 426.9 &  0.23 & 0.994 & 0.33 & 0.94 \\ 
J16525+4013 & 1.04$\times$0.51 & 12.9 & 63.7 & 50.7 &  0.23 & 0.991 & $-$0.14 & 0.80 \\ 
J16529+3902 & 1.06$\times$0.51 & 10.8 & 332.4 & 323.7 &  0.22 & 0.980 & $-$0.03 &  0.97 \\ 
J16542+3950 & 1.05$\times$0.52 & 7.4 & 77.7 & 57.2 &  0.21 & 0.987 & $-$0.38 & 0.74 \\ 
\enddata
\end{deluxetable}
\clearpage

\begin{deluxetable}{lrrrrrrrr}
\tabletypesize{\footnotesize}
\tablewidth{0pt}
\tablecolumns{9}
\tablecaption{Core Polarimetry Results\label{tab4}}
\tablehead{\colhead{Source} & 
\colhead{$P_{5}$} & \colhead{$m_{5}$} & \colhead{$\chi_{5}$} & 
\colhead{$P_{15}$} & \colhead{$m_{15}$} & \colhead{$\chi_{15}$} & 
\colhead{RM} & \colhead{$\chi_{B}$} \\
  \colhead{Name} & 
\colhead{(mJy beam$^{-1}$)} & \colhead{(\%)} &  \colhead{(deg)} & 
\colhead{(mJy beam$^{-1}$)} & \colhead{(\%)} &  \colhead{(deg)} & 
  \colhead{(rad m$^{-2}$)} & \colhead{(deg)} \\
\colhead{(1)}&  \colhead{(2)} & \colhead{(3)} & \colhead{(4)} & \colhead{(5)} &
\colhead{(6)} & \colhead{(7)} & \colhead{(8)} & \colhead{(9)} 
 }
\startdata
J08499+5108 & 3.6 & 3.9 & 24 & 0.5 & 0.5 & $-$77 & $-$282 $\pm$ 50 & 173 \\
J08546+5757 & 7.5 & 2.0 & 0 & 6.1 & 2.5 & 51 & $-$259 $\pm$ \phantom{0}7 & 147 \\
J09030+4651 & 1.3 & 0.8 & $-$18 & 1.7 & 0.8 & 48 & 585 $\pm$ 22 & 129 \\
J15450+5135 & 0.9 & 1.9 & 74 & 0.4 & 3.3 & $-$70 & $-$200 $\pm$ 60 & 28 \\
J15457+5400 & 1.8 & 1.2 & $-$18 & 3.9 & 3.2 & $-$26 & $-$931 $\pm$ 20 & 84 \\
J16469+4059 & 2.4 & 1.2 & 79 & 7.1 & 3.0 & 4 & $-$565 $\pm$ 14 & 106 \\
J16484+4104 & 1.2 & 0.6 & 81 & 1.4 & 0.4 & 89 & $-$22 $\pm$ 45 & 179 \\
J16529+3902 & 7.5 & 2.3 & $-$73 & 4.7 & 1.4 & 48 & 289 $\pm$ \phantom{0}8 & 136 \\
\enddata
\tablenotetext{~}{
Notes - (1) J2000 source name in the IAU format HHMMd+DDMM; (2) Linearly 
polarized flux density at 5 GHz; (3) Fractional linear polarization 
at 5 GHz; (4) Electric vector polarization angle at 5 GHz; 
(5) Linearly 
polarized flux density at 15 GHz; (6) Fractional linear polarization 
at 15 GHz; (7) Electric vector polarization angle at 15 GHz; 
(8) Faraday rotation measure; and (9) RM-corrected magnetic vector polarization 
angle.
}
\end{deluxetable}


\clearpage

\begin{thebibliography}{}
\bibitem[Abazajian et al.(2004)]{aba04} Abazajian, K., et 
al.\ 2004, AJ, 128, 502 

\bibitem[Abazajian et al.(2005)]{aba05} Abazajian, K., et 
al.\ 2005, AJ, submitted, astro-ph/0410239

\bibitem[Beasley et al.(2002)]{bea02} Beasley, A.~J., Gordon, 
D., Peck, A.~B., Petrov, L., MacMillan, D.~S., Fomalont, E.~B., \& Ma, C.\ 
2002, \apjs, 141, 13 

\bibitem[Condon et al.(1998)]{con98} Condon, J.~J., Cotton, 
W.~D., Greisen, E.~W., Yin, Q.~F., Perley, R.~A., Taylor, G.~B., \& 
Broderick, J.~J.\ 1998, AJ, 115, 1693 

\bibitem[Dermer \& Schlickeiser (1994)]{der94} Dermer, C.D. \& Schlickeiser, 
R. 1994, ApJS, 90 945

\bibitem[Falco, Kochanek, \& Mu\~noz (1998)]{fal98} Falco, E.~E., Kochanek, C.~S., \&
  Mu\~noz, J.~A. 1998, ApJ, 494, 47

\bibitem[Fanaroff \& Riley (1974)]{fan74} Fanaroff, B.~L. \& Riley,
  J.~M. 1974, MNRAS, 167, 31P


\bibitem[Gabuzda et al.(2000)]{gab00} Gabuzda, D.C. et al.\ 2000, 
MNRAS, 319, 1109

\bibitem[Gehrels \& Michelson(1999)]{geh99} Gehrels, N.~\& 
Michelson, P.\ 1999, Astroparticle Physics, 11, 277 

\bibitem[Gregory et al.(1996)]{gre96} Gregory, P.~C., Scott, 
W.~K., Douglas, K., \& Condon, J.~J.\ 1996, \apjs, 103, 427 

\bibitem[Greisen(2003)]{gre03} Greisen, E. W. 2003, in Information  
Handling in Astronomy - Historical Vistas, ed. A. Heck, Astrophysics
and Space Science Library Vol. 285 (Dordrecht: Kluwer), 109


\bibitem[Gugliucci et al.(2005)]{gug05} Gugliucci, N.~E., Taylor, G.~B., 
Peck, A.~B., \& Giroletti, M. 2005, ApJ, in press


\bibitem[Ma et al.(1998)]{ma98} Ma, C., et al.\ 1998, \aj, 
116, 516

\bibitem[Mattox et al.(1997)]{mat97} Mattox, J.~R., 
Schachter, J., Molnar, L., Hartman, R.~C., \& Patnaik, A.~R.\ 1997, \apj, 
481, 95 

\bibitem[Mattox et al.(2001)]{mat01} Mattox, J.~R., Hartman, 
R.~C., \& Reimer, O.\ 2001, \apjs, 135, 155 

\bibitem[Meier, Koide, \& Uchida (2001)]{mei01} Meier, D.L., 
Koide, S., \& Uchida, Y. 2001, Science, 291, 84

\bibitem[Myers et al.(2003)]{mye03} Myers, S.~T., et al.\ 
2003, MNRAS 341, 1 

\bibitem[Pearson \& Readhead (1988)]{pr88}
Pearson T. J., \& Readhead A. C. S.\ 1988, ApJ 328, 114

\bibitem[Peck \& Taylor(2000)]{pec00}Peck, A.~B. \& Taylor, G.~B. 2000, ApJ, 534, 90

\bibitem[Pollack et al.(2003)]{pol03}
Pollack L.K., Taylor G.B., \& Zavala R.T. 2003, ApJ 589, 733


\bibitem[Readhead et al.(1996)]{rea96} Readhead, A. C. S., Taylor,
G. B., Pearson, T. J., \& Wilkinson, P. N. 1996, ApJ, 460, 634

\bibitem[Shepherd(1997)]{she97} Shepherd, M.~C.\ 1997, ASP 
Conf.~Ser.~125: Astronomical Data Analysis Software and Systems VI, 6, 77 

\bibitem[Sowards-Emmerd et al.(2003)]{sow03} Sowards-Emmerd, 
D., Romani, R.~W., \& Michelson, P.~F.\ 2003, \apj, 590, 109 


\bibitem[Taylor et al.(1996)]{tay96} Taylor, G.~B., 
Vermeulen, R.~C., Readhead, A.~C.~S., Pearson, T.~J., Henstock, D.~R., \& 
Wilkinson, P.~N.\ 1996, ApJS, 107, 37 

\bibitem[Taylor(1998)]{tay98} Taylor, G.~B.\ 1998, \apj, 506, 637 

\bibitem[Taylor \& Myers (2000)]{tmy00} Taylor, G.~B. \& Myers, S.~T. 2000 
VLBA Scientific Memo 26, National Radio Astronomy Observatory

\bibitem[Taylor \& Peck (2003)]{tay03} Taylor, G.~B. \& Peck,
  A.~B. 2003, ApJ, 597, 157

\bibitem[Vercellone et al.(2004)]{ver04} Vercellone, S., 
Soldi, S., Chen, A.~W., \& Tavani, M.\ 2004, \mnras, 353, 890 

\bibitem[Weinstein et al.(2004)]{wei04} Weinstein, M.~A., et 
al.\ 2004, ApJS, 155, 243 

\bibitem[Zavala \& Taylor(2001)]{zav01} Zavala, R.~T.~\& 
Taylor, G.~B.\ 2001, \apjl, 550, L147 

\bibitem[Zavala \& Taylor (2003)]{zav03} Zavala, R.~T. \& Taylor,
  G.~B. 2003, \apj, 589, 126

\bibitem[Zavala \& Taylor (2004)]{zav04} Zavala, R.~T. \& Taylor,
  G.~B. 2004, \apj, in press, astro-ph/0405534

\end{thebibliography}
\end{document}